\documentclass[prd,twocolumn,superscriptaddress,showpacs,nofootinbib,preprintnumbers]{revtex4}

\usepackage{amsmath}

\usepackage{amsfonts}
\usepackage{graphicx}
\usepackage{dcolumn}
\usepackage{hyperref}
\def\bea{\begin{eqnarray}}
\def\eea{\end{eqnarray}}
\def\beqa{\begin{equation}}
\def\eeqa{\end{equation}}

\def\be{\begin{equation}}
\def\ee{\end{equation}}

\def\5{\overline 5}

\begin{document}

\title{Phantom damping of matter perturbations}

\author{Luca Amendola}

\affiliation{INAF/Osservatorio Astronomico di Roma, Viale Frascati 33, 00040 Monte
Porzio Catone (Roma), Italy}

\author{Shinji Tsujikawa}

\affiliation{Department of Physics, Gunma National College of Technology, Gunma
371-8530, Japan}

\author{M.~Sami}


\affiliation{IUCAA, Post Bag 4, Ganeshkhind, Pune 411 007, India}

\date{\today{}}

\begin{abstract}
Cosmological scaling solutions are particularly important in solving
the coincidence problem of dark energy. We derive the equations of
sub-Hubble linear matter perturbations for a general scalar-field
Lagrangian--including quintessence, tachyon, dilatonic ghost condensate
and k-essence--and solve them analytically for scaling solutions. 
We find that matter perturbations are always damped if a phantom field
is coupled to dark matter and identify the cases in which the gravitational 
potential is constant. This provides an interesting possibility
to place stringent observational constraints on scaling dark energy
models. 
\end{abstract}


\maketitle

\section{Introduction}

In the past years much efforts have been made to understand the nature
and the origin of dark energy$-$ a major puzzle of modern cosmology.
The accumulating observational data continues to confirm that about
70\% of the total energy density in the present universe corresponds
to unknown energy with a negative pressure \cite{obser}. A wide variety
of dark energy models have been proposed to address this problem
\cite{quint} (see Ref.~\cite{review} for review). For a viable
cosmological evolution the field energy density should remain sub-dominant
during the radiation/matter dominant era and become important only
at late times. Though the dynamically evolving scalar-field models
have an edge over the cosmological constant scenario, they too, in
general, are plagued with fine-tuning problems of initial conditions
and parameters of the models. Cosmological scaling solutions, in which
matter and dark energy follow the same background evolution, can alleviate
the so-called coincidence problem by providing a dynamical 
attractor \cite{CLW}. For a minimally coupled scalar field, however,
the energy density of the field decreases proportionally to that of
the background fluid for scaling solutions and hence the acceleration
of universe cannot be realized. Ordinary dark energy models are then
supplemented by additional features tuned to allow the exit from the
scaling regime at late times \cite{BCN}.

It has, however, been shown that if the coupling between dark energy
and dark matter is taken into account, one can achieve an accelerated
expansion in the scaling regime, thus alleviating the coincidence
problem \cite{coupled1,coupled3}. It is therefore important to be
able to distinguish between scaling and non-scaling solutions observationally.
It is already known that scaling solutions with accelerated expansion
give an acceptable fit to the supernovae (SN) Ia \cite{AGP}; therefore
the background behaviour seems to be insufficient in distinguishing
between a scaling universe and a non-scaling one. What is still lacking
is the investigation of the evolution of density perturbations in
such models, except for specific cases \cite{Amendola1,Amendola2}.
The goal of this paper is to find the perturbation equation in the
sub-Hubble regime for a very general Lagrangian and to solve it along
scaling solutions. Our analysis is applied to a wide variety of coupled
dark energy models including quintessence, tachyon, dilatonic ghost
condensate and k-essence. We will show that along scaling solutions
the equation of matter perturbations can be solved \emph{analytically}
in terms of observable background quantities (equation of state and
density parameter). In particular, the growth rate is found to be
unbounded, both from below and from above. Finally, we shall see that
when the field behaves as phantom (equation of state $w_{\phi }<-1$)
then linear matter perturbations are always damped. 
We call this phenomenon {\it phantom damping}.

The prototype of these stationary solutions is the standard coupled
scalar field with an exponential potential \cite{Amendola1}. Here,
however, the perturbations grow too fast due to the extra attraction
induced by the coupling and drive an unacceptable Integrated Sachs-Wolfe
(ISW) effect \cite{tva} on the Cosmic Microwave Background (CMB).
This problem is generally expected in accelerated scaling regimes,
both because of the stronger interaction and because the onset of
acceleration may occur quite earlier than usual. An intriguing result
in our paper is that the ISW effect vanishes for some parameter values.
Although these parameters are not observationally 
acceptable due to current supernovae constraints, 
this result shows that
the problem of an unacceptable ISW effect can
be alleviated by allowing the phantom field.

\section{Density perturbations and scaling solutions}

We start with the following general Lagrangian \cite{PT}
\begin{equation}
{\mathcal{S}}\, \, =\int d^{4}x\sqrt{-g}\left[\frac{R}{2}\, 
+p(X,\phi )\right]+{\mathcal{S}}_{m}[\phi ,g_{\mu \nu }]\, ,\label{lag}
\end{equation}
 where $X$ is the kinematic term of a scalar field $\phi $, i.e.,
$X\equiv -g^{\mu \nu }\partial _{\mu }\phi \partial _{\nu }\phi /2$.
Here $p$ is a scalar field Lagrangian which is the function of $X$
and $\phi $, and ${\mathcal{S}}_{m}$ is the action for matter fields
which are generally dependent on $\phi $.

The perturbation equations have been discussed in Refs.~\cite{Amendola1,Amendola2}
for the system of a coupled scalar field. Here we shall study the
evolution of matter perturbations for scaling solutions with a general
Lagrangian (\ref{lag}). Let us consider metric perturbations $\Psi $
and $\Phi $ in the longitudinal gauge about the flat 
Fridmann-Robertson-Walker (FRW) background: 
\begin{equation}
{\textrm{d}}s^{2}=-(1+2\Psi ){\textrm{d}}t^{2}+a^{2}(t)(1-2\Phi )
{\textrm{d}}x_{i}{\textrm{d}}x^{i}\, ,\label{eq:metric}
\end{equation}
 where $a(t)$ is a scale factor. One has $\Phi =\Psi $ in the absence
of anisotropic stress. We shall study a cosmological scenario in which
the Universe is filled by the field $\phi $ with an energy density
$\rho $ and by only one type of matter fluid with an energy density
$\rho _{m}$. We also assume that the field is coupled to matter fluid
with a coupling $Q$ defined by 
$Q\equiv -1/(\rho _{m}\sqrt{-g})\delta {\mathcal{S}}_{m}/\delta \phi $. 
In the general case one should insert also  radiation and baryons and leave 
them  uncoupled in order not to violate the  equivalence principle and other observations. 
Defining the matter density contrast $\delta _{m}\equiv \delta \rho _{m}/\rho _{m}$
and the dimensionless matter velocity divergence $\theta _{m}\equiv \nabla _{i}v_{(m)i}/H$,
where $H$ is the Hubble rate, we obtain the following perturbation
equations in Fourier space with wavelength $\lambda _{k}=Ha/k$ \cite{Amendola2}:
\begin{eqnarray}
\hspace*{-0.5em}\delta _{m}' & = & 
-\theta _{m}+3\Phi '+\sqrt{6}Q\varphi '\, ,\label{cdm-n1}\\
\hspace*{-0.5em} \theta _{m}' & = & 
-\left(2+\frac{H'}{H}+\sqrt{6}Qx\right)\theta _{m}
+\frac{1}{\lambda _{k}^{2}}(\Phi +\sqrt{6}Q\varphi ),\label{cdm-n2}
\end{eqnarray}
 where $\varphi \equiv \delta \phi /\sqrt{6}$ and $x\equiv \dot{\phi }/\sqrt{6}H$.
Here a prime denotes the derivative with respect to $N\equiv {\textrm{log}}\, (a)$.
In Newtonian limit the perturbation equation for the field $\phi $
takes the following form 
\begin{eqnarray}
\varphi ''+F(\phi )\varphi '+m(\phi )_{\textrm{eff}}^{2}\varphi 
+\frac{c_{s}^{2}}{\lambda _{k}^{2}}\varphi =
-\frac{\sqrt{6}c_{s}^{2}Q\Omega _{m}\delta _{m}}{2p_{X}}\, , 
&  & \label{dvarphi}
\end{eqnarray}
 where $p_{X}\equiv \partial p/\partial X$, $c_{s}^{2}\equiv p_{X}/\rho _{X}$
and $\Omega _{m}\equiv \rho _{m}/(3H^{2})$. $F(\phi )$ and $m(\phi )_{\textrm{eff}}$
are functions of $p(X,\phi )$.

The effective mass $m(\phi )_{\textrm{eff}}$ is expected to be negligible
if the field $\phi $ is responsible for dark energy. Then on sub-Hubble
scales the $(c_{s}^{2}/\lambda _{k}^{2})\varphi $ term is important,
ad its amplitude is forced to balance with the r.h.s. of Eq.~(\ref{dvarphi}).
Then we find 
\begin{equation}
\varphi \simeq -\frac{\sqrt{6}\lambda _{k}^{2}
Q\Omega _{m}\delta _{m}}{2p_{X}}\, .\label{eqphidelta}
\end{equation}
 The gravitational potential is expressed as \cite{Amendola2}
 \begin{equation}
\Phi =-\frac{3}{2}\lambda _{k}^{2}\left[(\delta _{m}
+3\lambda_k^{2}\theta _{m})\Omega _{m}+\delta _{\phi }
\Omega _{\phi }+6x\varphi p_{X}\right]\,.
\end{equation}
Since $\varphi $ is proportional to $\lambda_{k}^{2}$ on sub-Hubble
scales, $\Phi $ is approximately given by 
$\Phi \simeq -(3\lambda _{k}^{2}/2)\Omega _{m}\delta _{m}$.
Then by using Eqs.~(\ref{cdm-n1}), (\ref{cdm-n2}) and (\ref{eqphidelta}),
we find finally 
\begin{equation}
\delta _{m}''+\left(2+\frac{H'}{H}+\sqrt{6}Qx\right)\delta _{m}'
-\frac{3}{2}\Omega _{m}\left(1+2\frac{Q^{2}}{p_{X}}\right)\delta _{m}=0\,.
\label{eq:matter}
\end{equation}
 This is a very general equation which holds for any coupled scalar
field with Lagrangian $p(X,\phi )$, even when the coupling $Q$ depends
on the field $\phi $. In the general case it can be integrated numerically.
In the following we show that it may be integrated analytically for
a scaling cosmology in which the equation of state $w_{\phi }\equiv p/\rho $
and the energy ratio $\Omega _{\phi }\equiv \rho /(3H^{2})$ are constant.

In the flat FRW background one obtains
the following conservation equations 
\begin{eqnarray}
&  & \dot{\rho }+3H(1+w_{\phi })\rho =-Q\rho _{m}\dot{\phi }\, ,\label{geneeq1}\\
&  & \dot{\rho }_{m}+3H(1+w_{m})\rho _{m}=Q\rho _{m}\dot{\phi }\,,\label{geneeq2}
\end{eqnarray}
where $w_{m}\equiv p_{m}/\rho _{m}$. 
The equation for $H$ is given by 
\begin{equation}
\frac{\dot{H}}{H^{2}}=-\frac{3}{2}(1+w_{s})\, ,\label{eqhph}
\end{equation}
where the effective equation of state is defined as
$w_{s}\equiv (w_{\phi }\rho +w_{m}\rho _{m})/(\rho +\rho _{m})$.

Scaling solutions satisfy the condition $\rho \propto \rho _{m}$,
i.e., ${\textrm{d}}{\textrm{log}\,\rho }/{\textrm{d}}t
={\textrm{d}}{\textrm{log}\,\rho _{m}}/{\textrm{d}}t$.
Assuming that the coupling $Q$ is a constant in the scaling regime,
it was shown in Refs.~\cite{PT,TS} that the existence of scaling
solutions restricts the form of the scalar field Lagrangian to be
\begin{equation}
p(X,\phi )=X\, g\left(Xe^{\lambda \phi }\right)\, ,\label{pLag}
\end{equation}
where $g$ is any function in terms of $Y\equiv Xe^{\lambda \phi }$
and $\lambda $ is given by 
\begin{equation}
\lambda \equiv Q\frac{1+w_{m}-\Omega _{\phi }(w_{m}-w_{\phi })}
{\Omega _{\phi }(w_{m}-w_{\phi })}\, .\label{lQ}
\end{equation}
 For the Lagrangian (\ref{pLag}) we find that \cite{GNST}
\begin{equation}
\Omega _{\phi }=x^{2}(g+2g_{1})\, ,~~w_{\phi }
\Omega _{\phi }=x^{2}g\, ,\label{Omew}
\end{equation}
where $g_{n}\equiv Y^{n}g^{(n)}$.

In what follows we shall specialize the formula to the most relevant
case $w_{m}=0$, i.e., that of cold dark matter. Then by using Eqs.~(\ref{lQ})
and (\ref{Omew}), the effective equation of state is given by 
\begin{equation}
w_{s}=\Omega _{\phi }w_{\phi }=-\frac{Q}{Q+\lambda }\,.\label{eq:weff}
\end{equation}
 We note that this property holds irrespective of the form of the
function $g(Y)$. One has $w_{s}=0$ for $Q=0$ and $w_{s}\rightarrow -1$
in the limit $Q\gg \lambda >0$.

To discuss the fixed points of our system, it is convenient to introduce
two dimensionless quantities $x$ and $y$, 
defined by $x\equiv \dot{\phi }/(\sqrt{6}H)$
and $y\equiv e^{-\lambda \phi /2}/(\sqrt{3}H)$. 
Then the evolution equations (\ref{geneeq1}), (\ref{geneeq2}) and (\ref{eqhph}) for
the Lagrangian (\ref{pLag}) can be casted in the following autonomous
dynamical system: 
\begin{eqnarray}
\hspace*{-0.5em} x' & = & \frac{3}{2}x\left[1+x^{2}g-2A(g+g_{1})\right]\nonumber \\
 &  & +\frac{\sqrt{6}}{2}\left[A(Q+\lambda )(g+2g_{1})x^{2}-\lambda x^{2}-QA\right],\label{dx}\\
\hspace*{-0.5em} y' & = & -\frac{\sqrt{6}}{2}\lambda xy+\frac{3}{2}y(x^{2}g+1),\label{dy}
\end{eqnarray}
 where $A\equiv (g+5g_{1}+2g_{2})^{-1}$. We note that the equation
of state for the field $\phi $ reads \begin{equation}
w_{\phi }=-1+\frac{2x^{2}p_{X}}{\Omega _{\phi }}\, .\label{wphi}\end{equation}
 Therefore the field behaves as a phantom ($w_{\phi }<-1$) for $p_{X}<0$.

The ordinary (phantom) scalar field with an exponential potential
corresponds to the choice $g(Y)=\epsilon -c/Y$ (negative $\epsilon $
is a phantom). The dilatonic ghost condensate \cite{PT} and the tachyon
\cite{tach} also has scaling solutions, since the Lagrangians in
these models are written in the form (\ref{pLag}) by the choice $g(Y)=-1+cY$
and $g(Y)=-c\sqrt{1-2Y}/Y$, respectively. The critical points can
be found by setting $x'=y'=0$ in Eqs.~(\ref{dx}) and (\ref{dy}).
In fact the property of fixed points for coupled systems was discussed
in Ref.~\cite{GNST} for three classes of dark energy models mentioned
above.

There exists the following scaling solution for \textit{any} form
of the function $g(Y)$: \begin{eqnarray}
x=\frac{\sqrt{6}}{2(Q+\lambda )}\, , &  & \label{scalingx}
\end{eqnarray}
 as was shown in Eq.~(37) in Ref.~\cite{PT}. We note that $|Q+\lambda |>\sqrt{6}/2$
and that $g=-2Q(Q+\lambda )/3$ along the scaling solution. One can
easily check that this solution actually satisfies Eqs.~(\ref{dx})
and (\ref{dy}). We recall that $\Omega _{\phi }$ (and $g_{1}$)
remains undetermined and depends on the specific Lagrangian.

For the scaling solution (\ref{scalingx}) we have $Qx=-\sqrt{6}w_{s}/2$
and $p_{X}=g+g_{1}=(\Omega _{\phi }+w_{s})/(2x^{2})$ by Eqs.~(\ref{Omew})
and (\ref{eq:weff}). Then one can write the perturbation equation
(\ref{eq:matter}) in terms of $w_{s}$ and $\Omega _{\phi }$: 
\begin{equation}
\delta _{m}''+\xi _{1}\delta _{m}'+\xi _{2}\delta _{m}=0\,,
\end{equation}
where 
\begin{eqnarray}
 &  & \xi _{1}\equiv \frac{1}{2}-\frac{9}{2}w_{s}\, ,\\
 &  & \xi _{2}\equiv -\frac{3}{2}(1-\Omega _{\phi })
 \left(1+\frac{6w_{s}^{2}}{\Omega _{\phi }+w_{s}}\right)\, .
\end{eqnarray}
 Observationally, we expect $\Omega _{\phi }$ to be in the range
0.6-0.8 (from the complementary matter fraction in clustered objects)
and $w_{s}$ in the range $(-0.45,-0.75)$ from comparison with SNIa
\cite{AGP}.

\begin{figure}
\includegraphics[  bb=0bp 0bp 293bp 295bp,scale=0.6,clip]{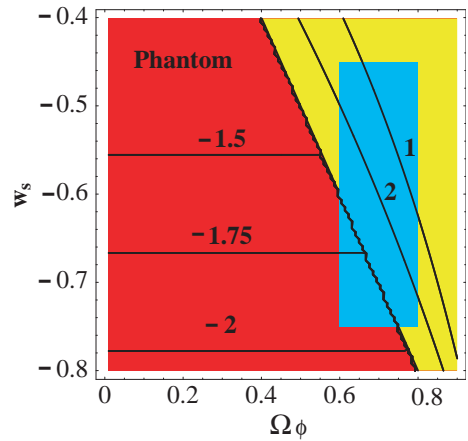}
\caption{Contour plot of the index $n_{+}$ in terms of the functions of $\Omega _{\phi }$
and $w_{s}$. The numbers which we show in the figure correspond to
the values $n_{+}$. In the non-phantom region characterized by $\Omega _{\phi }+w_{s}>0$,
$n_{+}$ are always positive. Meanwhile in the phantom region ($\Omega _{\phi }+w_{s}<0$)
with $w_{s}>-1$, $n_{+}$ take complex values with negative real
parts. We plot the real parts of $n_{+}$ in the phantom region. The
box (blue in the color version) represents schematically the observational
constraints on $w_{s},\Omega _{\phi }$ coming from the SNIa data. }
\label{fig1}
\end{figure}

Since $w_{s}$ and $\Omega _{\phi }$ are constants in the scaling
regime, we obtain the following solution 
\begin{equation}
\delta _{m}=c_{+}a^{n_{+}}+c_{-}a^{n_{-}}\, ,\label{delm}
\end{equation}
where $c_{\pm }$ are integration constants and 
\begin{equation}
n_{\pm }=\frac{1}{2}\left[-\xi _{1}\pm \sqrt{\xi _{1}^{2}-4\xi _{2}}\right]\,.
\label{power}
\end{equation}
One has $\Omega _{\phi }+w_{s}\equiv \Omega _{\phi }(1+w_{\phi })>0$
for a non-phantom scalar field, thus giving $n_{+}>0$ and $n_{-}<0$.
Therefore $\delta _{m}$ (and $\Phi $) grows in the scaling regime
($\delta _{m}\propto a^{n_{+}}$).

When $Q=0$ the solution of Eq.~(\ref{eq:matter}) for constant $w_{s}$
and $\Omega _{m}$ is given by Eq.~(\ref{delm}) with an index 
\begin{equation}
n_{\pm }=\frac{1}{2}\left[\frac{3}{2}w_{s}-\frac{1}{2}\pm 
\sqrt{\left(\frac{3}{2}w_{s}-\frac{1}{2}\right)^{2}+6\Omega _{m}}\right]\,.
\end{equation}
 Then we obtain $n_{+}=1$ and $n_{-}=-3/2$ in the matter dominant
era with $w_{s}\simeq 0$ and $\Omega _{m}\simeq 1$. From Eq.~(\ref{eq:weff})
one has $w_{s}=0$ in the uncoupled case ($Q=0$). Since $0\leq \Omega _{m}\leq 1$
in the scaling regime, the index $n_{+}$ satisfies $n_{+}\leq 1$
for uncoupled scaling solutions. Meanwhile the coupling $Q$ can lead
to the index $n_{+}$ larger than 1. In Fig.~1 we show
the contour plot of $n_{+}$ as the functions of $\Omega _{\phi }$
and $w_{s}$. The growth of the perturbations gets unboundedly larger
as we approach the border $\Omega _{\phi }+w_{s}=0$. The large index
$n_{+}$ obviously gives rise to a strong ISW effect on CMB, which
is not acceptable. However we caution that a precise bound on $n_{+}$
will depend on the specific choice of Lagrangian.

In the phantom region corresponding to $\Omega _{\phi }+w_{s}<0$,
we find that $n_{+}$ are complex with negative real parts for $w_{s}>-1$.
Therefore the perturbations exhibit damped oscillations in this case.
If $w_{s}<-1$, $n_{+}$ are either negative real values or complex
values with negative real parts. In any case the perturbations always
decay when the field $\phi $ corresponds to a phantom: we call this
phenomenon \emph{phantom damping}. In other words, the repulsive effect
of the phantom coupling dissipates the perturbations (at least in
the linear regime). 

Along this scaling solution the gravitational potential 
$\Phi \simeq -(3H^{2}a^2/2k^{2})\Omega _{m}\delta _{m}$
evolves as $a^{n_{+}-1-3w_{s}}$. It can be seen therefore that
the potential is constant for $n_{+}=3w_{s}+1$,  which
corresponds to $w_{s}^{\pm}=[-2 \pm \sqrt{4-3\Omega_{\phi}}]/3$.
Since $0 \le \Omega_{\phi} \le 1$ we find 
$-1/3 \le w_s^+ \le 0$ and $-4/3 \le w_s^- \le -1$
(for instance $w_s^+=-0.207$ and $w_s^-=-1.126$ for 
$\Omega _{\phi }=0.7$).
Although these values of $w_{s}$ are currently excluded 
by SN observations, it is interesting
to observe that there exist scaling solutions for which the 
gravitational potential is exactly constant.
This shows that, generally speaking, the absence of the ISW
effect does not imply the absence of dark energy.
We should also mention that values
of $w_s$ smaller than $-1$ are allowed if part of 
dark matter itself is not coupled, see Ref.~\cite{AGP}.
Clearly, the full investigations of such cases require 
the numerical integration of the Boltzmann equations
and it is beyond the scope of this paper.

\section{Conclusions}

In this paper we have studied the evolution of sub-Hubble linear perturbations
in the universe filled with a general scalar field coupled to dark
matter. We analytically derived the solutions for matter perturbations
when the background is described by the scaling solution given by
Eq.~(\ref{scalingx}). This analysis can be applied to any dark energy
models which possess scaling solutions.

The power-law index $n_{\pm }$ for perturbations is expressed by
the functions of $\Omega _{\phi }$ and $w_{s}$ only. The evolution
of perturbations is neatly divided by the border $\Omega _{\phi }+w_{s}=0$
between the ordinary field and the phantom field. While the perturbations
grow for $w_{\phi }>-1$, they are always suppressed in the phantom
case. In Fig.~\ref{fig1} we plot the index $n_{+}$ as the functions
of $\Omega _{\phi }$ and $w_{s}$ together with the constraint coming
from the SNIa datasets. In the non-phantom region the growth of the
perturbations gets larger as the parameters approach the border $\Omega _{\phi }+w_{s}=0$,
which would give rise to an unacceptable ISW effect. Therefore it
is likely that large part of the parameters space in Fig.~\ref{fig1}
is excluded from the CMB constraints. The phantom region allowed by
the SNIa constraint corresponds to the strong suppression 
with ${\rm Re}(n_{+})\lesssim -1.6$. 

It is certainly of interest to place constraints on $\Omega _{\phi }$
and $w_{s}$ using the latest CMB datasets. This requires a full detailed
analysis of the evolution of perturbations for each Fourier mode
without using the short wavelength approximation, which we leave to
future work. This will provide a powerful way to distinguish
between coupled dark energy models and other alternatives.

\begin{acknowledgments}
The work of S.~T. is supported by JSPS (No.\,30318802). M.~S. thanks Jamia~ Millia
for hospitality during the period of his leave from the university. 
\end{acknowledgments}

\end{document}